\documentclass[%
 aip,
 jmp,%
 amsmath,amssymb,
 reprint,%
]{revtex4-2}

\usepackage{graphicx}
\usepackage{dcolumn}
\usepackage{bm}

\newcommand{\wn}{cm$^{-1}$}
\newcommand{\chie}{\delta\chi_{\rm eff}}
\newcommand{\chieff}[1]{$\chie(#1)$}

\begin{document}

\preprint{AIP/123-QED}

\title[phase sensitive ISRS]{Low frequency coherent Raman imaging robust to optical scattering}



\author{David R. Smith}
\affiliation{Morgridge Institute for Research, Madison, WI} 

\author{Jesse W. Wilson}

\affiliation{Department of Electrical Engineering, Colorado State University, Fort Collins, CO}


\author{Siddarth Shivkumar}
\affiliation{Department of Physics, University of Ottawa, Ottawa, Ontario K1N6N5, Canada}

\author{Herv\'e Rigneault}

\affiliation{Aix Marseille Univ, CNRS, Centrale Med, Institut Fresnel, Marseille, France.}


\author{Randy A. Bartels}
\email{rbartels@morgride.org}
\affiliation{Morgridge Institute for Research, Madison, WI} 
\altaffiliation{Department of Biomedical Engineering, University of Wisconsin, Madison, WI}

\date{\today}

\begin{abstract}
We demonstrate low-frequency interferometric impulsive stimulated Raman scattering (ISRS) imaging with high robustness to distortions by optical scattering. ISRS is a pump-probe coherent Raman spectroscopy that can capture Raman vibrational spectra. Recording of ISRS spectra requires isolation of a probe pulse from the pump pulse. While this separation is simple in non-scattering specimens, such as liquids, scattering leads to significant pump pulse contamination and prevent the extraction of a Raman spectrum. We introduce a robust method for ISRS microscopy that works in complex scattering samples. High signal-to-noise ISRS spectra are obtained even when the pump and probe pulses pass through many scattering layers.
\end{abstract}

\maketitle

\section{Introduction}

\noindent Chemical imaging is a powerful imaging modality in the biological sciences. While fluorescent imaging with exogenous fluorescent probes or transgenically expressed fluorescent proteins forms the workhorse for chemical imaging in biological specimens \cite{Thorn:2016ef}, limitations of fluorescent imaging technologies have driven the development of label-free imaging modalities. \cite{TONG2011264, KASPROWICZ201789, Pavone, schie2013label, ijms22052657, 10.3389/fphy.2019.00170, schnell2020all} Of the many label-free imaging modalities, microscopy based on Raman scattering remains particularly appealing due to the narrow Raman vibrational spectroscopic lines that permit reliable identification of molecular species.

The vibrational frequency of materials and molecules offers a powerful means to both identify them and probe their dynamics and interactions. \cite{schie2013label} Spectral features of the vibrational spectrum can be optically interrogated through either direct dipole-allowed transitions with infrared Raman spectroscopy or through inelastic Raman scattering. Raman scattering is often preferred because infrared spectroscopy suffers from interference from water absorption, broad spectral features, and low spatial resolution resulting from the use of long optical wavelengths. \cite{schnell2020all} That said, while Raman microscopy with spontaneous scattering is common for many applications, the weak Raman scattering cross-section and incoherent nature of spontaneous Raman scattering severely limit the rate of scattered signal detection---leading to long integration times and poor sensitivity to low concentrations of Raman-active molecules. Raman signals can be enhanced by stimulating the Raman transitions, either by introducing near-field enhancements or by directly stimulating the transitions with a Stokes laser field. Coherent Raman scattering (CRS) \cite{TONG2011264} overcomes the weak Raman signals by invoking stimulated Raman scattering to drive a much larger response, enabling video-rate CRS microscopy. Moreover, an improved detection sensitivity for low concentration and weak Raman scattering modes is obtained with techniques that leverage homodyne interference, producing a linear dependence of the CRS signal on concentration \cite{TONG2011264, bartels2021low, HerveTutorial}.

Raman vibrational frequencies carry extremely valuable information for the study of molecules and material systems. Vibrational frequencies, $\Omega_v$, scale as a harmonic oscillator, $\Omega_v^2 = k/m$. Thus, strong binding, associated with a large value of the effective force constant $k$, and low (reduced) mass, $m$, of the vibrational mode correspond to high vibrational frequencies. The highest frequency modes are approximately localized and correspond to hydrogen bonds due to the low hydrogen mass and are used in investigations of water and lipids. The mid-band frequencies, $\Omega_v\sim400-1200$ \wn, are extremely powerful for the identification of particular molecules \cite{TONG2011264} and even for bacterial classification. \cite{BacterialClassifiation} Recent attention has focused on the largely neglected low-frequency vibrational modes. \cite{bartels2021low} These modes are generally associated with large reduced mass and correspond to vibrational motion that occurs over an extended region. Such relevant motions include virus capsid vibrations, \cite{tsen2007probing} deformations of proteins, \cite{ProteinLowFreqRaman} and mechanical properties of solids, particularly for soft \cite{LowFreqRamanSoftMat}  and two-dimensional materials. \cite{LowFreqRaman2DMat, LowFreqRama2DMatStack} 

However, implementation of Raman scattering at low vibrational frequencies remains a persistent challenge \cite{bartels2021low} because inelastic scattering at small offsets is difficult to measure, requiring low noise, narrow linewidth lasers, multiple stages of monochromators, and extremely steep spectral edge filters. The result is that imaging is extremely challenging. In this Letter, we introduce a simple and robust technique that is suitable for imaging in complex specimens. Our approach makes use of impulsive stimulated Raman scattering (ISRS) to probe low-frequency vibrational modes with time-domain spectroscopy.\cite{bartels2021low} In ISRS, a short pump pulse is used to drive the excitation of vibrational coherences for vibrational frequencies lower than $\sim 1/\tau_p$, where $\tau_p$ is the pump pulse duration. \cite{Dhar:1994sy}  Because the Raman interaction involves a change in molecular polarizability with vibrational displacement, excitation of vibrational coherences produces an effective time-varying perturbation to the linear optical susceptibility denoted by \chieff{t}. \cite{bartels2021low} The Raman spectrum is obtained by recording signals with a probe pulse that follows behind the pump pulse by a delay of $\tau$. The recorded signals are derived from the temporal phase, $\phi_v(t)$, accumulated by the probe pulse as it propagates through the excited vibrational coherence. This recovered spectrum is continuous and spans a range with a maximum frequency on the order of  $1/\tau_p$ down to a minimum frequency $1/\Delta\tau$ bounded by the range of pump-probe delay $\Delta\tau$.

The readout method of the Raman spectrum in ISRS impacts the sensitivity of spectral detection. \cite{bartels2021low} The most common technique is to make use of a spectral shift for detection, where spectral scattering from the time-dependent phase modulation applied to the probe pulse by the transient effective linear optical susceptibility leads to changes in power transmitted through a spectral filter. \cite{Merlin:1997ca, Domingue:2014bx,Raanan2019,smith2022nearly} Improved detection sensitivity can be achieved by turning the probe pulse spectral shift into a change in arrival time at a detector by applying additional spectral dispersion. \cite{smith2021phase} Methods that detect the signal based on a spectral shift reduce the amplitude of the low-frequency vibrational modes and impart a spectral filter on the relative amplitudes of the Raman spectrum. To emphasize the low-frequency vibrational modes and eliminate the spectral distortions imparted by frequency shift detection, \chieff{\tau} should be probed directly. \cite{bartels2021low}

Low frequency vibrations can be readily detected with ISRS methods where the signal detection is directly proportional to the time-varying change in the optical susceptibility induced by the forced Raman response. Several methods have been explored for the direct measurement of \chieff{t} to obtain an undistorted Raman spectrum, which we call phase-sensitive ISRS (ps-ISRS). ps-ISRS spectroscopy was originally implemented in a transient grating geometry, \cite{Dhar:1994sy} which allows for background-free measurements but relies on direct measurement of the diffracted beam power, leading to limited detection sensitivity. In the transient grating experiment, the amplitude of the diffracted probe field is proportional to \chieff{\tau}, i.e., the susceptibility perturbation at the pump-probe delay $\tau$. However, because the total scattered power is recorded, the Raman spectrum is distorted. The direct amplitude of \chieff{\tau} can be obtained with Kerr lensing \cite{Raanan:2018df} and interferometric measurements. \cite{Wahlstrand:05, Wilson:2008lk, Schlup:2009bq, Wilson:2008hh} A comparison of the Raman spectral amplitudes has been demonstrated in a Sagnac interferometer configuration and verified the access to low-frequency vibrational spectral information when \chieff{t} is directly obtained. \cite{Wahlstrand:05} The Sagnac interferometer configuration is intrinsically stable, but has limited application for spectroscopic or imaging measurements. A stable common path interferometer configuration is possible with a collinear time-delayed probe and reference pulse configuration, \cite{van2005detection, schlup2007dispersion} which has been demonstrated with spectral interferometry \cite{Wilson:2008lk, Schlup:2009bq, Wilson:2008hh} and Fourier transform interferometry. \cite{ghosh2021broadband}

\begin{figure}[htbp]
\centering
\includegraphics[width=\linewidth]{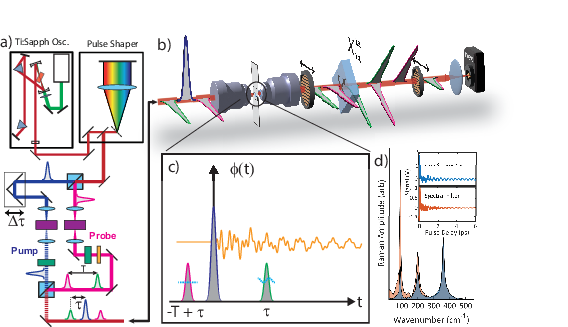}
\caption{Common path interferometric ps-ISRS concept figure. a) Experimental setup for ps-ISRS. The acousto-optic modulators, calcite crystals, and linear polarizer are represented by the purple, green, and orange elements, respectively. b) Conceptual figure that shows the interaction of the pump, probe, and reference pulses with the sample, as well as the isolation of the probe-reference pair and re-timing to produce pulse interference. The interference produces signal current changes in a photodiode that are proportional to the Raman-induced transient phase, $\phi(\tau)$, at the arrival time, $\tau$, of the probe pulse. c) The relative arrival time of the pump, probe, and reference pulses. d) Example Raman spectra measured in BGO for ps-ISRS and spectral shift detection \cite{Bartels:2001wo}. The improved low-frequency Raman detection with ps-ISRS is evident.}
\label{fig:InterferometricConcept}
\end{figure}

\section{Concept}

Unfortunately, the existing ps-ISRS methods are restricted to non-depolarizing and non-scattering samples such as liquid-phase spectroscopy or transparent crystals.  In this Letter, we introduce a simple collinear interferometric approach to ps-ISRS that is extremely stable and thus performs robustly in the presence of strong scattering and depolarization effects, such as in cell culture and tissues.

A conceptual schematic of the experimental system is shown in Fig. \ref{fig:InterferometricConcept}. A modelocked ultrafast laser produces pulses that are passed through a spatial light modulator pulse shaper for second and third order dispersion control. The laser beam power is split into pump and probe pulses in an orthogonally polarized Mach-Zenhder interferometer. The pump pulse intensity is modulated by an acousto-optic modulator to allow for lock-in amplifier detection of the Raman signal. A computer-controlled resonant delay scanner in the pump arm enables rapid scanning of the relative pump-probe delay $\tau$. The probe pulse arm is modified from a standard pump-probe experimental arrangement \cite{Domingue:2014bx, Gershgoren:2003tx,smith2022nearly} in order to generate a probe-reference pulse pair with a birefringent optical crystal. \cite{schlup2007dispersion} 

Birefringent crystals are used both as a pulse pair generator \cite{schlup2007dispersion, wilson2011rapid, Domingue:2014bx} and a pulse retimer to create a common path collinearly propagating probe-reference pulse pair with an extremely stable relative phase and a temporal separation of $T$, as shown in Fig. \ref{fig:InterferometricConcept}b. The probe-reference pair is generated by sending a linearly polarized probe pulse through a birefringent crystal plate rotated so that the eigen-polarization axes are at 45 degrees with respect to the input linear polarization of the probe pulse. Depending on the cut of the crystal and the orientation of the optic axis, the relative delay time, $T$, can be adjusted by rotating the birefringent crystal or by adjusting the crystal thickness. The pulses are separated by a few picoseconds and share a common propagation path. 

To combine the probe-reference pulse pair with the pump pulse, the probe-reference pair is projected onto the same linear polarization axis in the polarizing beam splitter where the pulse pair is combined with the orthogonally polarized pump pulse. The pulses are focused into the sample with a high numerical aperture aspheric lens and recollimated with an identical aspheric lens after passing through the specimen. The pump pulse is removed from the collimated beams with a polarizer to isolate the probe-reference pulse pair. 

The coherent ISRS signal is contained in a phase perturbation, $\phi_v(t)$, induced by the pump pulse. The reference pulse precedes the pump pulse, accumulating a phase as it propagates through the optical system. This phase accumulated by the reference pulse comes from the sample at equilibrium because the vibrational coherences decay to equilibrium (over several picoseconds), long before the next pulse in the oscillator pulse train arrives ($\sim 10$ ns). By contrast, the probe pulse arrives at a delay, $\tau$, after the pump pulse and accumulates a phase identical to the reference pulse in addition to a phase perturbation induced by Raman excitation by the pump pulse. By placing the pump pulse in between the reference and probe pulses, the relative phase between the probe and reference pulses is the accumulated vibrational phase perturbation acquired by the probe pulse. After interacting with the sample, the probe-reference pair is isolated from the pump pulse---for example with a polarizer---then the probe-reference pulse pair is re-timed. In this way, the relative phase is converted to an amplitude modulation through interference between the probe and reference pulse so that the Raman signal is detected with a simple photodiode. 

This pulse pair is re-timed using a second birefringent crystal that is oriented at --45 degrees to undo the probe-reference delay so that the probe and reference pulses overlap in time, i.e., setting $T=0$. Because the probe-reference pulse pair propagates along the same direction and are only separated by a few picoseconds, $T\sim 6$ ps, the relative phase between these two pulses is exceptionally stable---allowing very stable phase measurements of the transient phase introduced by the pump pulse. The probe and reference pulses are in the same spatial mode and polarization state; only half of the power of the probe and reference pulses is overlapped in time and will interfere (see Fig. \ref{fig:InterferometricConcept} b)). In addition, the two pulses travel through exactly the same optical path, and thus share the same spatial mode, making interference between the beams robust to perturbations acquired with propagation through optical scattering environments. To demonstrate the robustness of our approach, we introduce controlled scattering to a sample by adding layers of parafilm on the sample between the objective and the sample of interest. For a more detailed description of the experimental setup, please refer to the supplementary information. 

The interferometric signal for ISRS spectroscopy emerges from the fact that the pump pulse excites a non-equilibrium time-varying change in effective linear optical susceptibility, \chieff{t}, imparting a transient phase modulation on a time-delayed probe pulse.  The reference pulse can be arranged to arrive before the pump pulse so that the only phase that it accumulates is from the optical system and the specimen at thermal equilibrium. The pump pulse-induced phase perturbation, $\phi_v(t) =  k_{\rm pr} \, \ell \, \chie(t) /2 n_0 $, is then the only phase shift acquired between the probe and reference pulses. Here, $\ell$ is the focal interaction length, $n_0$ is the sample refractive index at thermal equilibrium, and $k_{\rm pr} = 2 \, \pi /\lambda$ is the free-space wavenumber for a pulse with center wavelength $\lambda$. The fact that the probe and reference pulses are common path means that all accumulated scattering, aberrations, and depolarizations are identical for both beams. This produces an extremely stable signal that prevents many forms of technical noise.

\begin{figure}[htbp]
\centering
\includegraphics[width=\linewidth]{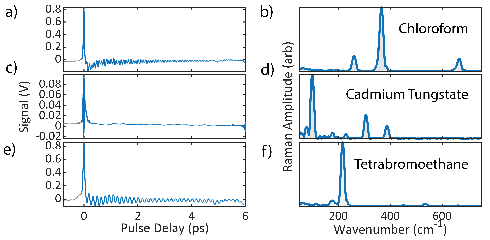}
\caption{ps-ISRS data from multiple samples with low frequency Raman vibrations. The left column shows the time-resolved signal for each sample and the right column shows the corresponding Raman spectrum.}
\label{fig:multiSample}
\end{figure}

The phase modulation due to the excited vibrational coherence can be expressed as a linear superposition of the phase modulation from each excited vibrational mode as $\delta \phi^v(\tau) = \sum_v  \delta \phi_0^v \ h_v(\tau)$, and depends on the arrival time $\tau$ of the probe pulse after the excitation induced by the pump and is summed over the contribution of each vibrational mode, denoted with a superscript $v$. Each vibrational mode is excited through an impulsive forced Raman response that is driven by the pump pulse, leading to a damped causal oscillation \cite{HerveTutorial,bartels2021low}, $h_v(t) = \Theta(t) \,  \exp( - \Gamma_v  \ t  /2) \, \sin(\Omega_v t)$. Here $\Theta(t)$ is the Heavyside step function that ensures causality and $\Gamma_v $ is the damping rate of the mode at the vibrational frequency $\Omega_v$. The induced vibrational phase perturbation $\delta \phi_0^v = g_f \,  \mathrm{Im}[\chi^{(3)}_{\mathrm VR}(\Omega_v)] \,  \Gamma_v \, \tilde{D}(\Omega_v)  \, \bar{p}_{\rm pu}$ scales linearly with the average power of the pump pulse, $\bar{p}_{\rm pu}$, and the molecular concentration, $N$, which becomes clear when we note that $\mathrm{Im}[\chi^{(3)}_{\mathrm VR}(\Omega_v)] = N \, (\partial \alpha/\partial Q)^2_0/6 \epsilon_0 \, \Gamma_v \, \Omega_v$. The amplitude of the Raman-induced phase modulation is also proportional to the power spectral density (PSD) of the pulse intensity profile  at the vibrational frequency $\Omega_v$ \cite{Bartels:2001wo,Bartels:2002le}, $\tilde{D}(\Omega) = \mathcal{F} \{ I(t) \}/  I_0 \, \tau_0$. Here $I_0$ is the peak pulse intensity and $\tau_0 = \int I(t)/I_0 \, d t$ is a measure of the pulse duration. In the limit of an impulsive pump pulse, $I(t) = I_0 \, \delta(t)$, then the excitation PSD, $\tilde{D}(\Omega) \rightarrow 1$, so that the forced Raman response is the impulse for vibrational excitation. Here, we assume that the temporal duration, $\tau_p$, and the temporal intensity, $I(t)$, of the pump, probe, and reference pulses are all identical. The coherent Raman excitation that drives the vibrational phase perturbation also depends on the focusing conditions, which are encapsulated in the focusing parameter $g_f = 12 \, \pi  \, \ell/(n^2 \, \epsilon_0 \, c \, \nu_R \, \mathcal{A}_f \, \lambda)$, where $\mathcal{A}_f$ is the focal beam cross sectional area, $\nu_R$ is the laser repetition rate frequency, and $\lambda$ is probe pulse center wavelength.

\begin{figure}[htbp]
\centering
\includegraphics[width=\linewidth]{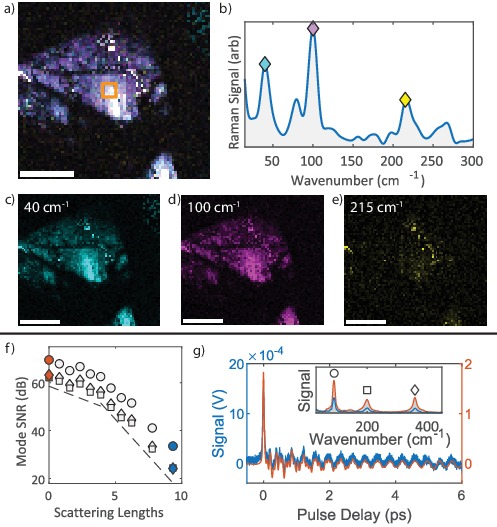   }
\caption{a) A composite image of the three low-frequency Raman-active optical phonon modes in an anthracene crystal. b) The Raman spectrum of the crystal at a single pixel centered in the orange square shown in a). c--e) Individual images of the three Raman modes indicated in b). f) The signal-to-noise ratio (SNR) of the three Raman vibrational modes of a BGO crystal with scattering layers added between the sample and the excitation objective. A high SNR is indicated for $> 10$ scattering lengths. g) A comparison of the Raman spectrum of BGO for zero and 12 scattering layers shows high quality spectra extracted even under conditions of strong optical scattering.}
\label{fig:scattering}
\end{figure}

\section{Results}

In our experiment, the probe-reference pulse pair, with a total average power of $\bar{p}_{\rm pr}$, is isolated from the pump pulse using polarizer to block the pump pulse. Then, half of the probe and reference pulse average power is adjusted to interfere in a compact, collinear interferometer using an angle-tuned birefringent crystal. The relative phase of probe and reference pulses are adjusted to be in quadrature so that the total power of the interfering pulses is $\bar{p}_{\rm int} = \bar{p}_{\rm pr} \left\{ 1+ \sin[\delta \phi^v(\tau)]\right\} /2$ , leading to a signal power incident on the detector of $\bar{p}_{\rm  sig}(\tau) = \delta\phi^v(\tau) \ \bar{p}_{\rm pr}/2$, with a background power equal to the total average probe pulse train power $\bar{p}_{\rm  bkg} = \bar{p}_{\rm pr}$. In the shot noise limit, the signal-to-noise ratio (SNR) for a detection integration time of $\Delta t$ reads $\mathrm{SNR} = (\delta\phi^v/2) \ \Upsilon \ \sqrt{\Delta t}$, where $\Upsilon = \sqrt{\mathcal{R} \ \bar{p}_{\rm pr} / e}$, $\mathcal{R}$ is the detector responsivity, and $e$ is the elementary charge. Provided that the total probe power exceeds approximately 2 mW, the shot noise limit is valid for lock-in detection that is able to largely avoid the laser intensity fluctuation noise. Because both the signal power and SNR depend linearly on $\delta \phi^v$, this collinear interferometric method is sensitive to low Raman vibrational frequencies, and is limited where the impulsive Raman excitation strength, $\tilde{D}(\Omega_v)$, becomes too low, which occurs when the vibrational period is on the order or larger than the pump pulse duration.

The ps-ISRS spectrum of the low-frequency optical phonon Raman spectrum in BGO is shown in Fig. \ref{fig:scattering}d in comparison to a spectral shift measurement. The increased amplitude at low vibrational frequencies is apparent in the interferometric measurement. Because the pump-probe delay $\tau$ can be controlled independently of the separation between the probe and reference pulses, which is a fixed delay $T$, there are three zones for the ps-ISRS experimental system, each of which is illustrated in Fig. S2. We define the time-frame so that $t=0$ occurs at the center of the pump pulse The first pulse to arrive is denoted the reference pulse that is incident at at a time $t=\tau-T$ and the probe pulse arrives at a time $\tau$. Zone I occurs when $\tau<0$, so that both the probe and reference pulses arrive before the pump pulse. In this region, there is no pump-induced transient phase acquired between the probe and reference pulses, so we have a null signal. Zone II is where we pull the Raman spectra; this temporal signal provides a pure Raman spectrum because only the probe pulse accumulated the Raman-induced phase shift. In Zone III, the spectrum is distorted because the interference signal is determined by the phase shift acquired by the probe and reference pulses, i.e., $\phi(\tau) - \phi(T-\tau)$.

Inteferometric ps-ISRS spectra from several samples displaying low-frequency Raman modes are shown in Fig. \ref{fig:multiSample}. While the chloroform and tetrabromoethane are liquid samples and do not suffer depolarization, the cadmium tungstate is a birefringent crystal and induces a depolarization between the pump pulse and the probe-reference pulse pair. The time-domain traces are shown on the left column, whereas the Raman spectra shown in the right column is computed from the power spectral density of the time-domain trace after isolation of the signal that contains only the Raman transients away from the cross phase modulation feature near time zero. \cite{Wilson:2008hh} The observed Raman spectral modes match known Raman spectral lines of these three samples. 

Hyperspectral Raman images of anthracene crystals for three low-frequency Raman-active optical phonon modes are shown in Fig. \ref{fig:scattering}. The Raman spectrum of the optical phonon modes is shown in Fig. \ref{fig:scattering}b. Images for the three phonon modes marked in the spectrum are shown in Fig. \ref{fig:scattering}c-e, whereas a composite image is displayed in Fig. \ref{fig:scattering}a. The complex morphology of the crystals produces scattering that distorts the beams and degrades the Raman signal. However, we see that high-quality images are obtained with the common-path interferometric readout because the probe and reference pulses maintain identical spatial and polarization modes.

To quantitatively evaluate the effect of optical scattering on ps-ISRS with common-path interferometric detection, we captured Raman spectra as we added layers of optical scattering material. For each measurement, the Raman signal and the noise were estimated from the temporal trace, from which we computed the SNR. To introduce optical scattering, layers of parafilm were added on the front of the BGO sample, thus introducing scattering before Raman excitation by the pump pulse and spectral scattering of the probe pulse. Results are shown in Fig. \ref{fig:scattering}f, displaying the SNR as a function of the number of scattering layers added between the sample and the excitation objective. The data show two regions of SNR decay as a function of the number of scattering lengths, indicated by dashed lines in the figure. The Raman spectra for the highest and lowest SNR data are shown in Fig. \ref{fig:scattering}g.

\section{Discussion}

Imaging of low-frequency vibrational modes with coherent Raman remains a challenging prospect. Here we introduce a new imaging strategy for a wide bandwidth of Raman vibrational frequencies, spanning from very low frequencies ($<300$ cm$^{-1}$) to frequencies well within the fingerprint region $>1000$ cm$^{-1}$. Our new strategy relies on a pump-probe geometry using ps-ISRS excitation of a vibrational coherence by a short pump pulse that is electronically non-resonant with the specimen. Such short pulse excitation is capable of exciting vibrational frequencies with periods that are longer than the pump pulse duration. \cite{Bartels:2001wo} As a result, short laser pulses with carefully maintained dispersion compensation are critical for ISRS excitation. After excitation of the vibrational coherence, the readout of the signal with a probe pulse plays a crucial role in the vibrational frequencies that are observed. The conventional strategy is to use a spectral edge filter to convert the spectral scattering driven by the transient phase modulation acquired by the probe pulse when propagating through the vibrational coherence into a change in probe pulse power transmitted through the spectral filter. \cite{Domingue:2014bx, Ren:19} Unfortunately, this simple strategy renders ISRS largely insensitive to low-frequency vibrations. \cite{bartels2021low} By reading out the phase directly through interferometry, we directly detect the time domain-forced Raman response, providing sensitive detection of the low-frequency Raman vibrational modes. 

Aside from the high sensitivity to low-frequency vibrational modes (e.g., terahertz Raman), we have demonstrated that our common-mode strategy for the detection of the transient Raman phase modulation produces high SNR data even in the face of optical scattering. This resilience of our signal detection to optical scattering follows from two properties that we acquire by generating the probe-reference pair in a suitably oriented birefringent crystal. The first property is that the probe and reference pulses are common path. Because of this common path propagation, any scattering that is encountered is common to both the probe and reference pulse beams. As a result, the beams are guaranteed to still interfere. Indeed, even depolarization is not a problem, as the spatial modes of the probe and reference beams are identical---ensuring stable interference even after becoming severely distorted, as they are distorted identically.

Of course, common path perturbations are only relevant if the relative phase, scattering, and depolarization are stable for the duration of the integration time over which the interference is acquired. The relative stability of the probe and reference pulses is guaranteed by the fact that there is very little perturbation of the refractive index of the specimen that can occur on the timescale of a few picoseconds other than the non-equilibrium vibrational coherence dynamics that are excited with the pump pulse. 

The combination of the temporal and common-mode interferometric stability ensures extremely stable and high-sensitivity detection of the vibration-induced phase shift. The robustness of our approach is demonstrated by introducing controlled scattering to a sample by adding layers of parafilm on the sample between the objective and the sample of interest.  The scattering introduced by the parafilm layers was characterized by measuring power transmission through a blank sample as more layers of parafilm were added. The power transmitted through the scattering layers decreases exponentially with the number of scattering layers.  The scattering length, defined as when the optical power decreases by $1/e$, was found to be $\ell_s = 1.29$ parafilm layers.  Figure \ref{fig:scattering}f shows the SNR for the three main vibrational modes of BGO as a function of scattering length of parafilm added to the sample. Panel (g) of Figure \ref{fig:scattering} shows the time-resolved signal and the Raman spectrum (inset) of BGO with no scattering layers (orange) and the signal with the maximum number of scattering layers used (blue), which is 12 layers of parafilm, or 9.34 scattering lengths.  This result shows the robustness of the interferometric Raman technique, with the SNR of the BGO modes remaining high after 9.34 scattering lengths.  The incident optical powers at the sample were $p_{pu} = 37.8$ mW and $p_{pr} = 38.95$ mW for the pump and probe, respectively.  After passing through the sample, objectives, retiming crystal and analyzing polarizer, the probe power incident on the detection photodiode was $p_{pr} = 8.6$ mW with no scattering present.  The probe power incident on the detection photodiode was measured to be $p_{pr} = 7.7$ $\mu$W with 12 parafilm layers (9.34 scattering lengths).

Because we have constructed this system based on orthogonally polarized pump and probe-reference pulses, we are limited to vibrational modes with a more symmetric character, and we may have reduced sensitivity to asymmetric modes. The spectral resolution of this strategy is limited to $\sim 1/T$ for methods that rely on power spectral estimation, e.g. the FFT or multitaper computation. However, the use of model-based estimation, e.g. LP-SVD, can circumvent this spectral resolution limitation. The lower bound of the vibrational frequency is limited by the scan range of the probe-reference pair. For Zome II, the lower bound is $\tilde{\nu} = (c \, T)^{-1} \sim 7$ cm$^{-1}$, but this can be extended arbitrarily with synthetic aperture scanning \cite{Wilson:2008lk}, where the probe and reference pulse pair both follow the pump pulse, at the cost of the loss of frequencies with vibrational periods that are an integer multiple of $T$. Application of LD-SVD methods are expected to further improve this lower bound. 

\section{Conclusions}

We have developed a new and versatile method for low-frequency coherent Raman imaging. Balanced detection could eliminate the need for the acousto-optic modulator for beam intensity modulation, which will drastically reduce the spectral dispersion burden of the system. Such an approach would allow for easier use of high numerical aperture objective lenses that also carry a high dispersion. Right now, the relatively low numerical aperture and off-axis aberrations of the aspheres have limited our ability to image in complex samples, and the long working distance makes thin samples such as two-dimensional materials challenging to image. Future advances will explore additional strategies to simplify the imaging approach.

\begin{acknowledgments}
We are grateful for the help of Dekel Raanan and Maor Asher for crystal synthesis as well as Dan Oron for helpful discussions and assistance with sample. We are grateful for funding from DE-FOA-0002603 (Bartels). Alicia Williams from the Morgridge Institute provided scientific editing support.  ANR-21-ESRS-0002 IDEC, EU ICT 101016923 CRIMSON, EU EIC 101099058 VIRUSONG, EU ERC 101052911 (Rigneault). NIH 5 R21GM135772-02 (Wilson and Bartels)
\end{acknowledgments}

\appendix

\section{Appendixes}

\section{ps-ISRS experimental system}

\begin{figure*}[htbp]
    \begin{center}
        \resizebox{0.95\linewidth}{!}{\includegraphics[width = 1.5\textwidth]{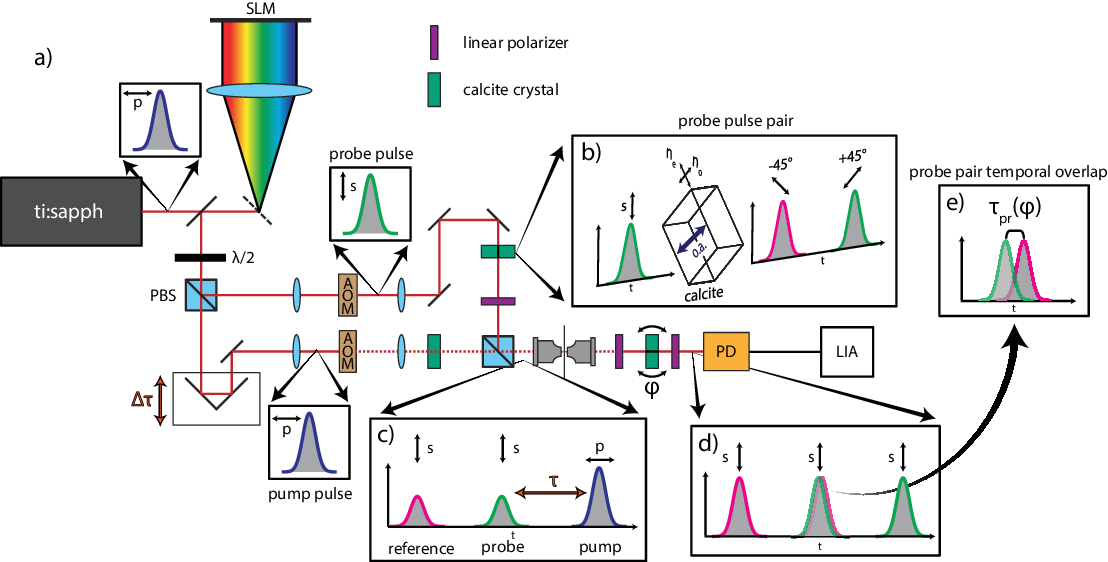}}
        \caption{Experimental layout. a) Pump-probe experimental layout with an ultrafast pulse laser, and ultrafast pulse shaper, and a orthogonal, linearly polarized Mach-Zehnder interferometer. b) Illustration of the oriented uniaxial crystal that is used to produce a very stable probe-reference pulse pair from a single probe pulse. The pulse pair never propagates in free space ans are only delayed by a few picoseconds. Thus, the relative phase stability is extremely robust. c) When the pump pulse and probe-reference pulse pair are recombined on the output polarizing beam splitter, the probe-reference pulses are projected onto the same linear polarization direction that is perpendicular the the linearly polarized pump pulse direction. d) After the probe and reference pulses are re-timed, we have a set of three pulses, where the middle pulse comes from the interference of the probe and reference pulses and the two satellite pulses are the residual pulses projected onto the analyzer polarization direction.}
        \label{fig:ExpLayout}
    \end{center}
\end{figure*}

The experimental layout for the system is provided in Fig. \ref{fig:ExpLayout}. The key to a robust ps-ISRS imaging is the production of common-path probe and reference pulses.\cite{van2005detection,  schlup2007dispersion, Hartinger:2008oe} This pulse pair is generated in the experimental system that builds on a pump-probe experimental system using a Mach-Zehnder interferometer.\cite{smith2022nearly} In addition, because we are using an ISRS interaction with broad-bandwidth ultrafast pulses, careful dispersion management is required. Pulses are generated in a Ti:sapphire laser (KM Labs Halcyon, $\sim ~ 600$ mW average power, $\sim 18$ fs pulses, 94 MHz repetition rate). The output laser beam is collimated and then directed into a pulse shaper constructed from a folded Martinez stretcher with a flat mirror and a transmission spatial light modulator (SLM) placed at the focal plane of the lens (the Martinez Fourier plane). \cite{domingue2015nearly, JesseCalibrattion} Details of the pulse compression protocol can be found in the supplemental information in our recent split-spectrum ISRS paper. \cite{smith2022nearly}

After passing through the pulse shaper, the beam is split into pump and probe (reference) pulses with a low-dispersion polarizing beam splitter (ATFilms, PBS1005-GVD). The relative power in the two arms is controlled by a half-wave plate placed before the beam splitter. An acousto-optic modulator (AOM) crystal resides in each arm of the pump-probe interferometer. In the pump arm, the pump pulse is passed through an AOM that imparts a sinusoidal intensity modulation onto the pump pulse train for lock-in detection. The pump pulses are modulated at several MHZ to allow for detection of the signal through lock-in detection at an offset frequency where intensity noise fluctuations from the laser are low. The AOM in probe arm is left passive and is required to balance the spectral dispersion in the two interferometer arms. The pump and probe pulses are recombined with a second low-dispersion polarizing beam splitter. A reflective optical delay line in the pump pulse arm is used to control the relative arrival time, $\tau$, of the pump and probe pulses. The pump-probe delay is rapidly scanned with a resonant scanner delay line (Peregrine, Mesa Photonics) that is capable of performing 20 ps delay scans at 60 Hz. The full delay is scanned with a triangle pattern that produces full ps-ISRS signal traces at a rate of 120 Hz. A full Raman spectrum is obtained in $\sim$8.3 ms. The Raman spectrum is obtained from the temporal trace by either estimating the spectrum of the temporal trace with a fast Fourier transform through linear prediction singular value decomposition. \cite{Wilson:2008lk, wilson2011rapid, Sinjab:20}

To enable ps-ISRS detection, the probe pulses are temporally split into a pair of orthogonally-polarized probe and reference pulses. The probe and reference pulses are created by sending the linearly polarized probe pulse into the birefringent crystal (10mm thick calcite, Newlight Photonics) with the crystal rotated so that the axis of the linear polarization of the probe pulse is at 45 degrees with respect to the eigen-polarization axes of the ordinary and extraordinary polarization directions. \cite{van2005detection, schlup2007dispersion, Hartinger:2008oe, Wilson:2008lk, JesseCalibrattion, Schlup:2009bq, Domingue:2014bx, ghosh2021broadband} The group velocity of the pulses that are projected onto the ordinary and extraordinary polarization directions differ, leading time delay accumulated through propagation through the uniaxial birefringent crystal. The rotation angle of the optical axis of the crystal can be adjusted to fine-tune the relative delay and the relative phase between the probe and reference pulses. Then the probe-reference pair is combined with the pump pulse in the polarizing beam splitter, the pulses are projected on the same linear polarization with equal amplitude. The pump pulse is orthogonally polarized to the probe-reference pair. An identical birefringent crystal is placed in pump arm to mach the dispersion between the arms.

The combined and orthogonally polarized pump and probe-reference pulses are focused into the with a low-dispersion aspheric lens (40x/0.6 NA, Newport).  Typical pump/probe powers at the focus are $p_{pu} = 38\textrm{ mW}$, $p_{pr} = 19.5 \textrm{ mW}$, and $p_{ref} = 19.5 \textrm{ mW}$. After passing through the sample, the beams are collimated with a matching asphere. Once the probe-reference pulse pair is isolated with a polarizer, the time-delay is collapsed with a birefringent crystal oriented in the opposite direction of the original probe-reference pulse pair generator. The relative delay between the pulses is set to zero by adjusting the crystal angle to tune the relative delay to zero. The crystal angle is fine-tuned to adjust the relative phase so to $\pi/2$ or $-\pi/2$ to provide optimal sensitivity to the small phase shift accumulated by the probe pulse. After passing through an analyzing polarizer, the pulses interfere and convert the phase shift accumulated by the probe pulse to an amplitude modulation of the pulse interference. Note that because the probe and reference pulses were polarized along the same direction, each pulse is projected onto the ordinary and extraordinary polarization directions, so that half of the field amplitude for each pulse is further delayed in time. This is illustrated in Fig. \ref{fig:ExpLayout}d). The ps-ISRS signal is detected by an amplified photodiode (Thorlabs, PDA36A2)  that is directed into a lock-in amplifier (Zurich Instruments, HF2LI).

\begin{figure*}[htbp]
    \begin{center}
        \resizebox{0.95\linewidth}{!}{\includegraphics[width = 1.5\textwidth]{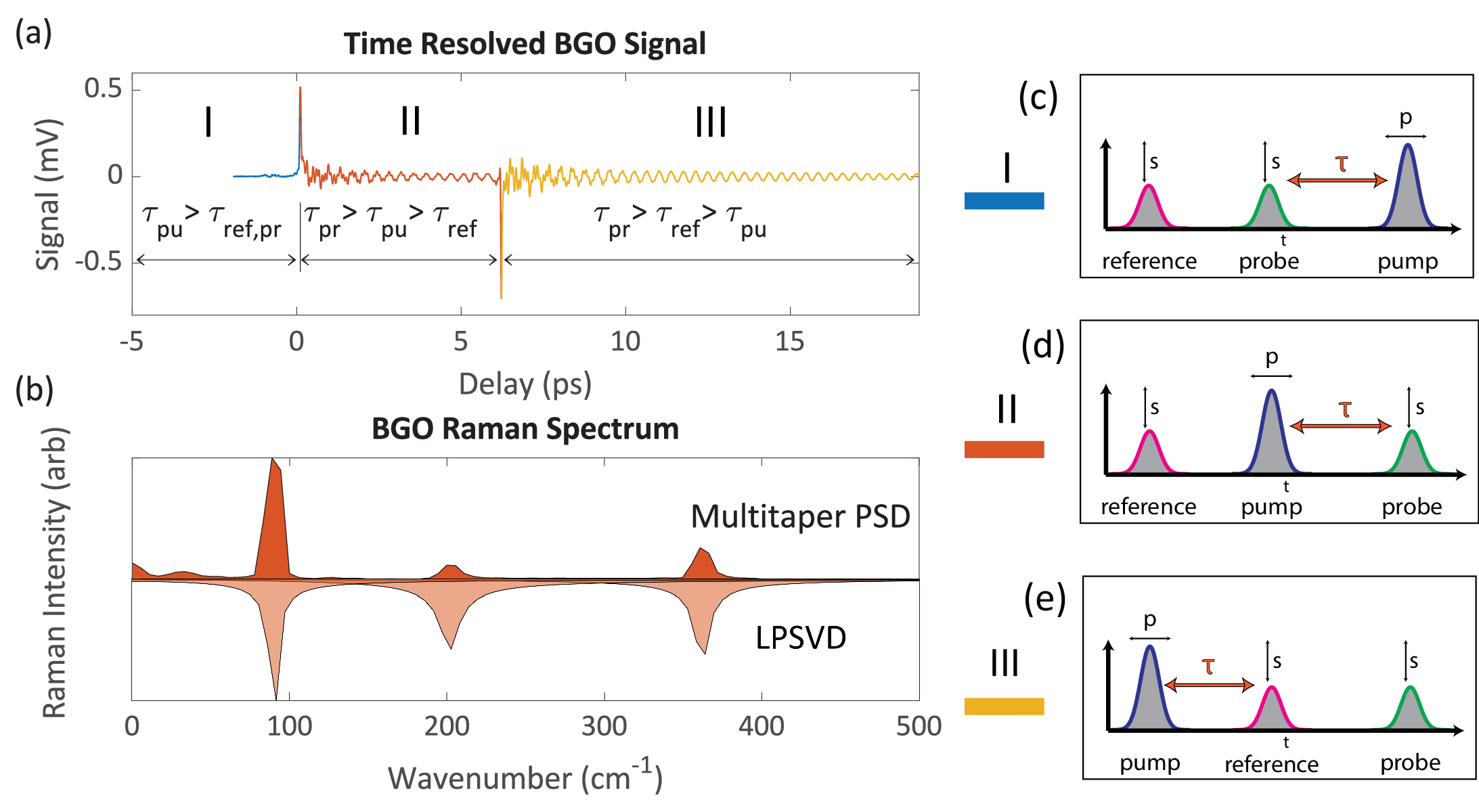}}
        \caption{a) Three zones of probe-reference pulse delays pump probe interferometry. In zone I, the probe and the reference pulses arrive before the pump pulse. Thus the probe and reference beams experience the same thermal equilibrium phase, so there is zero phase difference between these pulses, and thus no signal. In zone II, the reference pulse arrives before the pump pulse. Next the pump pulse arrives and excites the Raman vibrational coherence that are proved by the probe pulse. Thus, the phase difference that produces the signal is from non-equilibrium phase perturbation induced by the pump pulse. Finally, in zone III, both the reference pulse and the probe pulse arrive after the pump pulse. The phase difference depends on the difference in the non-equilibrium phase probed by reference and probe pulses. b) The power spectrum estimated from Zone II gives the Raman spectrum. The probe and reference pulses are linearly polarized along the same direction and perpendicular to the linear pump polarization. c) Illustrates Zone I where both probe and reference pulses arrive before the pump pulse. d) Illustrates Zone II where the pump pulse arrives in-between the probe and reference pulses. e) Illustrates Zone IIII where the probe and references pulses arrive after the pump pulse.}
        \label{fig:expPulseOrder}
    \end{center}
\end{figure*}

\bibliography{psISRS}

\providecommand{\noopsort}[1]{}\providecommand{\singleletter}[1]{#1}%
\begin{thebibliography}{39}%
\makeatletter
\providecommand \@ifxundefined [1]{%
 \@ifx{#1\undefined}
}%
\providecommand \@ifnum [1]{%
 \ifnum #1\expandafter \@firstoftwo
 \else \expandafter \@secondoftwo
 \fi
}%
\providecommand \@ifx [1]{%
 \ifx #1\expandafter \@firstoftwo
 \else \expandafter \@secondoftwo
 \fi
}%
\providecommand \natexlab [1]{#1}%
\providecommand \enquote  [1]{``#1''}%
\providecommand \bibnamefont  [1]{#1}%
\providecommand \bibfnamefont [1]{#1}%
\providecommand \citenamefont [1]{#1}%
\providecommand \href@noop [0]{\@secondoftwo}%
\providecommand \href [0]{\begingroup \@sanitize@url \@href}%
\providecommand \@href[1]{\@@startlink{#1}\@@href}%
\providecommand \@@href[1]{\endgroup#1\@@endlink}%
\providecommand \@sanitize@url [0]{\catcode `\\12\catcode `\$12\catcode `\&12\catcode `\#12\catcode `\^12\catcode `\_12\catcode `\%12\relax}%
\providecommand \@@startlink[1]{}%
\providecommand \@@endlink[0]{}%
\providecommand \url  [0]{\begingroup\@sanitize@url \@url }%
\providecommand \@url [1]{\endgroup\@href {#1}{\urlprefix }}%
\providecommand \urlprefix  [0]{URL }%
\providecommand \Eprint [0]{\href }%
\providecommand \doibase [0]{https://doi.org/}%
\providecommand \selectlanguage [0]{\@gobble}%
\providecommand \bibinfo  [0]{\@secondoftwo}%
\providecommand \bibfield  [0]{\@secondoftwo}%
\providecommand \translation [1]{[#1]}%
\providecommand \BibitemOpen [0]{}%
\providecommand \bibitemStop [0]{}%
\providecommand \bibitemNoStop [0]{.\EOS\space}%
\providecommand \EOS [0]{\spacefactor3000\relax}%
\providecommand \BibitemShut  [1]{\csname bibitem#1\endcsname}%
\let\auto@bib@innerbib\@empty
\bibitem [{\citenamefont {Thorn}\ and\ \citenamefont {Kellogg}(2016)}]{Thorn:2016ef}%
  \BibitemOpen
  \bibfield  {author} {\bibinfo {author} {\bibfnamefont {K.}~\bibnamefont {Thorn}}\ and\ \bibinfo {author} {\bibfnamefont {D.}~\bibnamefont {Kellogg}},\ }\bibfield  {title} {\enquote {\bibinfo {title} {A quick guide to light microscopy in cell biology},}\ }\bibfield  {booktitle} {\emph {\bibinfo {booktitle} {Molecular Biology of the Cell}},\ }\href {https://doi.org/10.1091/mbc.e15-02-0088} {\bibfield  {journal} {\bibinfo  {journal} {Molecular Biology of the Cell}\ }\textbf {\bibinfo {volume} {27}},\ \bibinfo {pages} {219--222} (\bibinfo {year} {2016})}\BibitemShut {NoStop}%
\bibitem [{\citenamefont {Tong}\ and\ \citenamefont {Cheng}(2011)}]{TONG2011264}%
  \BibitemOpen
  \bibfield  {author} {\bibinfo {author} {\bibfnamefont {L.}~\bibnamefont {Tong}}\ and\ \bibinfo {author} {\bibfnamefont {J.-X.}\ \bibnamefont {Cheng}},\ }\bibfield  {title} {\enquote {\bibinfo {title} {Label-free imaging through nonlinear optical signals},}\ }\href {https://doi.org/https://doi.org/10.1016/S1369-7021(11)70141-9} {\bibfield  {journal} {\bibinfo  {journal} {Materials Today}\ }\textbf {\bibinfo {volume} {14}},\ \bibinfo {pages} {264--273} (\bibinfo {year} {2011})}\BibitemShut {NoStop}%
\bibitem [{\citenamefont {Kasprowicz}, \citenamefont {Suman},\ and\ \citenamefont {O'Toole}(2017)}]{KASPROWICZ201789}%
  \BibitemOpen
  \bibfield  {author} {\bibinfo {author} {\bibfnamefont {R.}~\bibnamefont {Kasprowicz}}, \bibinfo {author} {\bibfnamefont {R.}~\bibnamefont {Suman}},\ and\ \bibinfo {author} {\bibfnamefont {P.}~\bibnamefont {O'Toole}},\ }\bibfield  {title} {\enquote {\bibinfo {title} {Characterising live cell behaviour: Traditional label-free and quantitative phase imaging approaches},}\ }\href {https://doi.org/https://doi.org/10.1016/j.biocel.2017.01.004} {\bibfield  {journal} {\bibinfo  {journal} {The International Journal of Biochemistry \& Cell Biology}\ }\textbf {\bibinfo {volume} {84}},\ \bibinfo {pages} {89--95} (\bibinfo {year} {2017})}\BibitemShut {NoStop}%
\bibitem [{\citenamefont {Cicchi}\ and\ \citenamefont {Pavone}(2014)}]{Pavone}%
  \BibitemOpen
  \bibfield  {author} {\bibinfo {author} {\bibfnamefont {R.}~\bibnamefont {Cicchi}}\ and\ \bibinfo {author} {\bibfnamefont {F.~S.}\ \bibnamefont {Pavone}},\ }\bibfield  {title} {\enquote {\bibinfo {title} {Multimodal nonlinear microscopy: A powerful label-free method for supporting standard diagnostics on biological tissues},}\ }\href {https://doi.org/10.1142/S1793545813300085} {\bibfield  {journal} {\bibinfo  {journal} {Journal of Innovative Optical Health Sciences}\ }\textbf {\bibinfo {volume} {07}},\ \bibinfo {pages} {1330008} (\bibinfo {year} {2014})},\ \Eprint {https://arxiv.org/abs/https://doi.org/10.1142/S1793545813300085} {https://doi.org/10.1142/S1793545813300085} \BibitemShut {NoStop}%
\bibitem [{\citenamefont {Schie}\ and\ \citenamefont {Huser}(2013)}]{schie2013label}%
  \BibitemOpen
  \bibfield  {author} {\bibinfo {author} {\bibfnamefont {I.~W.}\ \bibnamefont {Schie}}\ and\ \bibinfo {author} {\bibfnamefont {T.}~\bibnamefont {Huser}},\ }\bibfield  {title} {\enquote {\bibinfo {title} {Label-free analysis of cellular biochemistry by raman spectroscopy and microscopy},}\ }\href@noop {} {\bibfield  {journal} {\bibinfo  {journal} {Comprehensive Physiology}\ }\textbf {\bibinfo {volume} {3}},\ \bibinfo {pages} {941--956} (\bibinfo {year} {2013})}\BibitemShut {NoStop}%
\bibitem [{\citenamefont {Borile}\ \emph {et~al.}(2021)\citenamefont {Borile}, \citenamefont {Sandrin}, \citenamefont {Filippi}, \citenamefont {Anderson},\ and\ \citenamefont {Romanato}}]{ijms22052657}%
  \BibitemOpen
  \bibfield  {author} {\bibinfo {author} {\bibfnamefont {G.}~\bibnamefont {Borile}}, \bibinfo {author} {\bibfnamefont {D.}~\bibnamefont {Sandrin}}, \bibinfo {author} {\bibfnamefont {A.}~\bibnamefont {Filippi}}, \bibinfo {author} {\bibfnamefont {K.~I.}\ \bibnamefont {Anderson}},\ and\ \bibinfo {author} {\bibfnamefont {F.}~\bibnamefont {Romanato}},\ }\bibfield  {title} {\enquote {\bibinfo {title} {Label-free multiphoton microscopy: Much more than fancy images},}\ }\href {https://doi.org/10.3390/ijms22052657} {\bibfield  {journal} {\bibinfo  {journal} {International Journal of Molecular Sciences}\ }\textbf {\bibinfo {volume} {22}} (\bibinfo {year} {2021}),\ 10.3390/ijms22052657}\BibitemShut {NoStop}%
\bibitem [{\citenamefont {Mazumder}\ \emph {et~al.}(2019)\citenamefont {Mazumder}, \citenamefont {Balla}, \citenamefont {Zhuo}, \citenamefont {Kistenev}, \citenamefont {Kumar}, \citenamefont {Kao}, \citenamefont {Brasselet}, \citenamefont {Nikolaev},\ and\ \citenamefont {Krivova}}]{10.3389/fphy.2019.00170}%
  \BibitemOpen
  \bibfield  {author} {\bibinfo {author} {\bibfnamefont {N.}~\bibnamefont {Mazumder}}, \bibinfo {author} {\bibfnamefont {N.~K.}\ \bibnamefont {Balla}}, \bibinfo {author} {\bibfnamefont {G.-Y.}\ \bibnamefont {Zhuo}}, \bibinfo {author} {\bibfnamefont {Y.~V.}\ \bibnamefont {Kistenev}}, \bibinfo {author} {\bibfnamefont {R.}~\bibnamefont {Kumar}}, \bibinfo {author} {\bibfnamefont {F.-J.}\ \bibnamefont {Kao}}, \bibinfo {author} {\bibfnamefont {S.}~\bibnamefont {Brasselet}}, \bibinfo {author} {\bibfnamefont {V.~V.}\ \bibnamefont {Nikolaev}},\ and\ \bibinfo {author} {\bibfnamefont {N.~A.}\ \bibnamefont {Krivova}},\ }\bibfield  {title} {\enquote {\bibinfo {title} {Label-free non-linear multimodal optical microscopy---basics, development, and applications},}\ }\href {https://doi.org/10.3389/fphy.2019.00170} {\bibfield  {journal} {\bibinfo  {journal} {Frontiers in Physics}\ }\textbf {\bibinfo {volume} {7}} (\bibinfo {year} {2019}),\ 10.3389/fphy.2019.00170}\BibitemShut {NoStop}%
\bibitem [{\citenamefont {Schnell}\ \emph {et~al.}(2020)\citenamefont {Schnell}, \citenamefont {Mittal}, \citenamefont {Falahkheirkhah}, \citenamefont {Mittal}, \citenamefont {Yeh}, \citenamefont {Kenkel}, \citenamefont {Kajdacsy-Balla}, \citenamefont {Carney},\ and\ \citenamefont {Bhargava}}]{schnell2020all}%
  \BibitemOpen
  \bibfield  {author} {\bibinfo {author} {\bibfnamefont {M.}~\bibnamefont {Schnell}}, \bibinfo {author} {\bibfnamefont {S.}~\bibnamefont {Mittal}}, \bibinfo {author} {\bibfnamefont {K.}~\bibnamefont {Falahkheirkhah}}, \bibinfo {author} {\bibfnamefont {A.}~\bibnamefont {Mittal}}, \bibinfo {author} {\bibfnamefont {K.}~\bibnamefont {Yeh}}, \bibinfo {author} {\bibfnamefont {S.}~\bibnamefont {Kenkel}}, \bibinfo {author} {\bibfnamefont {A.}~\bibnamefont {Kajdacsy-Balla}}, \bibinfo {author} {\bibfnamefont {P.~S.}\ \bibnamefont {Carney}},\ and\ \bibinfo {author} {\bibfnamefont {R.}~\bibnamefont {Bhargava}},\ }\bibfield  {title} {\enquote {\bibinfo {title} {All-digital histopathology by infrared-optical hybrid microscopy},}\ }\href@noop {} {\bibfield  {journal} {\bibinfo  {journal} {Proceedings of the National Academy of Sciences}\ }\textbf {\bibinfo {volume} {117}},\ \bibinfo {pages} {3388--3396} (\bibinfo {year} {2020})}\BibitemShut {NoStop}%
\bibitem [{\citenamefont {Bartels}, \citenamefont {Oron},\ and\ \citenamefont {Rigneault}(2021)}]{bartels2021low}%
  \BibitemOpen
  \bibfield  {author} {\bibinfo {author} {\bibfnamefont {R.~A.}\ \bibnamefont {Bartels}}, \bibinfo {author} {\bibfnamefont {D.}~\bibnamefont {Oron}},\ and\ \bibinfo {author} {\bibfnamefont {H.}~\bibnamefont {Rigneault}},\ }\bibfield  {title} {\enquote {\bibinfo {title} {Low frequency coherent raman spectroscopy},}\ }\href@noop {} {\bibfield  {journal} {\bibinfo  {journal} {Journal of Physics: Photonics}\ }\textbf {\bibinfo {volume} {3}},\ \bibinfo {pages} {042004} (\bibinfo {year} {2021})}\BibitemShut {NoStop}%
\bibitem [{\citenamefont {Rigneault}\ and\ \citenamefont {Berto}(2018)}]{HerveTutorial}%
  \BibitemOpen
  \bibfield  {author} {\bibinfo {author} {\bibfnamefont {H.}~\bibnamefont {Rigneault}}\ and\ \bibinfo {author} {\bibfnamefont {P.}~\bibnamefont {Berto}},\ }\bibfield  {title} {\enquote {\bibinfo {title} {Tutorial: Coherent raman light matter interaction processes},}\ }\href {https://doi.org/10.1063/1.5030335} {\bibfield  {journal} {\bibinfo  {journal} {APL Photonics}\ }\textbf {\bibinfo {volume} {3}},\ \bibinfo {pages} {091101} (\bibinfo {year} {2018})},\ \Eprint {https://arxiv.org/abs/https://doi.org/10.1063/1.5030335} {https://doi.org/10.1063/1.5030335} \BibitemShut {NoStop}%
\bibitem [{\citenamefont {Strola}\ \emph {et~al.}(2014)\citenamefont {Strola}, \citenamefont {Baritaux}, \citenamefont {Schultz}, \citenamefont {Simon}, \citenamefont {Allier}, \citenamefont {Espagnon}, \citenamefont {Jary},\ and\ \citenamefont {Dinten}}]{BacterialClassifiation}%
  \BibitemOpen
  \bibfield  {author} {\bibinfo {author} {\bibfnamefont {S.~A.}\ \bibnamefont {Strola}}, \bibinfo {author} {\bibfnamefont {J.-C.}\ \bibnamefont {Baritaux}}, \bibinfo {author} {\bibfnamefont {E.}~\bibnamefont {Schultz}}, \bibinfo {author} {\bibfnamefont {A.~C.}\ \bibnamefont {Simon}}, \bibinfo {author} {\bibfnamefont {C.}~\bibnamefont {Allier}}, \bibinfo {author} {\bibfnamefont {I.}~\bibnamefont {Espagnon}}, \bibinfo {author} {\bibfnamefont {D.}~\bibnamefont {Jary}},\ and\ \bibinfo {author} {\bibfnamefont {J.-M.}\ \bibnamefont {Dinten}},\ }\bibfield  {title} {\enquote {\bibinfo {title} {{Single bacteria identification by Raman spectroscopy}},}\ }\href {https://doi.org/10.1117/1.JBO.19.11.111610} {\bibfield  {journal} {\bibinfo  {journal} {Journal of Biomedical Optics}\ }\textbf {\bibinfo {volume} {19}},\ \bibinfo {pages} {111610} (\bibinfo {year} {2014})}\BibitemShut {NoStop}%
\bibitem [{\citenamefont {Tsen}\ \emph {et~al.}(2007)\citenamefont {Tsen}, \citenamefont {Dykeman}, \citenamefont {Sankey}, \citenamefont {Tsen}, \citenamefont {Lin},\ and\ \citenamefont {Kiang}}]{tsen2007probing}%
  \BibitemOpen
  \bibfield  {author} {\bibinfo {author} {\bibfnamefont {K.-T.}\ \bibnamefont {Tsen}}, \bibinfo {author} {\bibfnamefont {E.~C.}\ \bibnamefont {Dykeman}}, \bibinfo {author} {\bibfnamefont {O.~F.}\ \bibnamefont {Sankey}}, \bibinfo {author} {\bibfnamefont {S.-W.~D.}\ \bibnamefont {Tsen}}, \bibinfo {author} {\bibfnamefont {N.-T.}\ \bibnamefont {Lin}},\ and\ \bibinfo {author} {\bibfnamefont {J.~G.}\ \bibnamefont {Kiang}},\ }\bibfield  {title} {\enquote {\bibinfo {title} {Probing the low-frequency vibrational modes of viruses with raman scattering—bacteriophage m13 in water},}\ }\href@noop {} {\bibfield  {journal} {\bibinfo  {journal} {Journal of biomedical optics}\ }\textbf {\bibinfo {volume} {12}},\ \bibinfo {pages} {024009} (\bibinfo {year} {2007})}\BibitemShut {NoStop}%
\bibitem [{\citenamefont {Rischel}\ \emph {et~al.}(1998)\citenamefont {Rischel}, \citenamefont {Spiedel}, \citenamefont {Ridge}, \citenamefont {Jones}, \citenamefont {Breton}, \citenamefont {Lambry}, \citenamefont {Martin},\ and\ \citenamefont {Vos}}]{ProteinLowFreqRaman}%
  \BibitemOpen
  \bibfield  {author} {\bibinfo {author} {\bibfnamefont {C.}~\bibnamefont {Rischel}}, \bibinfo {author} {\bibfnamefont {D.}~\bibnamefont {Spiedel}}, \bibinfo {author} {\bibfnamefont {J.~P.}\ \bibnamefont {Ridge}}, \bibinfo {author} {\bibfnamefont {M.~R.}\ \bibnamefont {Jones}}, \bibinfo {author} {\bibfnamefont {J.}~\bibnamefont {Breton}}, \bibinfo {author} {\bibfnamefont {J.-C.}\ \bibnamefont {Lambry}}, \bibinfo {author} {\bibfnamefont {J.-L.}\ \bibnamefont {Martin}},\ and\ \bibinfo {author} {\bibfnamefont {M.~H.}\ \bibnamefont {Vos}},\ }\bibfield  {title} {\enquote {\bibinfo {title} {Low frequency vibrational modes in proteins: Changes induced by point-mutations in the protein-cofactor matrix of bacterial reaction centers},}\ }\href {https://doi.org/10.1073/pnas.95.21.12306} {\bibfield  {journal} {\bibinfo  {journal} {Proceedings of the National Academy of Sciences}\ }\textbf {\bibinfo {volume} {95}},\ \bibinfo {pages} {12306--12311} (\bibinfo {year} {1998})},\ \Eprint
  {https://arxiv.org/abs/https://www.pnas.org/doi/pdf/10.1073/pnas.95.21.12306} {https://www.pnas.org/doi/pdf/10.1073/pnas.95.21.12306} \BibitemShut {NoStop}%
\bibitem [{\citenamefont {Amin}\ \emph {et~al.}(2014)\citenamefont {Amin}, \citenamefont {Blake}, \citenamefont {Kenyon}, \citenamefont {Kennel},\ and\ \citenamefont {Lewis}}]{LowFreqRamanSoftMat}%
  \BibitemOpen
  \bibfield  {author} {\bibinfo {author} {\bibfnamefont {S.}~\bibnamefont {Amin}}, \bibinfo {author} {\bibfnamefont {S.}~\bibnamefont {Blake}}, \bibinfo {author} {\bibfnamefont {S.~M.}\ \bibnamefont {Kenyon}}, \bibinfo {author} {\bibfnamefont {R.~C.}\ \bibnamefont {Kennel}},\ and\ \bibinfo {author} {\bibfnamefont {E.~N.}\ \bibnamefont {Lewis}},\ }\bibfield  {title} {\enquote {\bibinfo {title} {A novel combination of dls-optical microrheology and low frequency raman spectroscopy to reveal underlying biopolymer self-assembly and gelation mechanisms},}\ }\href {https://doi.org/10.1063/1.4903785} {\bibfield  {journal} {\bibinfo  {journal} {The Journal of Chemical Physics}\ }\textbf {\bibinfo {volume} {141}},\ \bibinfo {pages} {234201} (\bibinfo {year} {2014})},\ \Eprint {https://arxiv.org/abs/https://doi.org/10.1063/1.4903785} {https://doi.org/10.1063/1.4903785} \BibitemShut {NoStop}%
\bibitem [{\citenamefont {Sriv}, \citenamefont {Kim},\ and\ \citenamefont {Cheong}(2018)}]{LowFreqRaman2DMat}%
  \BibitemOpen
  \bibfield  {author} {\bibinfo {author} {\bibfnamefont {T.}~\bibnamefont {Sriv}}, \bibinfo {author} {\bibfnamefont {K.}~\bibnamefont {Kim}},\ and\ \bibinfo {author} {\bibfnamefont {H.}~\bibnamefont {Cheong}},\ }\bibfield  {title} {\enquote {\bibinfo {title} {Low-frequency raman spectroscopy of few-layer 2h-sns2},}\ }\href {https://doi.org/10.1038/s41598-018-28569-6} {\bibfield  {journal} {\bibinfo  {journal} {Scientific Reports}\ }\textbf {\bibinfo {volume} {8}},\ \bibinfo {pages} {10194} (\bibinfo {year} {2018})}\BibitemShut {NoStop}%
\bibitem [{\citenamefont {Puretzky}\ \emph {et~al.}(2015)\citenamefont {Puretzky}, \citenamefont {Liang}, \citenamefont {Li}, \citenamefont {Xiao}, \citenamefont {Wang}, \citenamefont {Mahjouri-Samani}, \citenamefont {Basile}, \citenamefont {Idrobo}, \citenamefont {Sumpter}, \citenamefont {Meunier},\ and\ \citenamefont {Geohegan}}]{LowFreqRama2DMatStack}%
  \BibitemOpen
  \bibfield  {author} {\bibinfo {author} {\bibfnamefont {A.~A.}\ \bibnamefont {Puretzky}}, \bibinfo {author} {\bibfnamefont {L.}~\bibnamefont {Liang}}, \bibinfo {author} {\bibfnamefont {X.}~\bibnamefont {Li}}, \bibinfo {author} {\bibfnamefont {K.}~\bibnamefont {Xiao}}, \bibinfo {author} {\bibfnamefont {K.}~\bibnamefont {Wang}}, \bibinfo {author} {\bibfnamefont {M.}~\bibnamefont {Mahjouri-Samani}}, \bibinfo {author} {\bibfnamefont {L.}~\bibnamefont {Basile}}, \bibinfo {author} {\bibfnamefont {J.~C.}\ \bibnamefont {Idrobo}}, \bibinfo {author} {\bibfnamefont {B.~G.}\ \bibnamefont {Sumpter}}, \bibinfo {author} {\bibfnamefont {V.}~\bibnamefont {Meunier}},\ and\ \bibinfo {author} {\bibfnamefont {D.~B.}\ \bibnamefont {Geohegan}},\ }\bibfield  {title} {\enquote {\bibinfo {title} {Low-frequency raman fingerprints of two-dimensional metal dichalcogenide layer stacking configurations},}\ }\bibfield  {booktitle} {\emph {\bibinfo {booktitle} {ACS Nano}},\ }\href {https://doi.org/10.1021/acsnano.5b01884} {\bibfield
  {journal} {\bibinfo  {journal} {ACS Nano}\ }\textbf {\bibinfo {volume} {9}},\ \bibinfo {pages} {6333--6342} (\bibinfo {year} {2015})}\BibitemShut {NoStop}%
\bibitem [{\citenamefont {Dhar}, \citenamefont {Rogers},\ and\ \citenamefont {Nelson}(1994)}]{Dhar:1994sy}%
  \BibitemOpen
  \bibfield  {author} {\bibinfo {author} {\bibfnamefont {L.}~\bibnamefont {Dhar}}, \bibinfo {author} {\bibfnamefont {J.~A.}\ \bibnamefont {Rogers}},\ and\ \bibinfo {author} {\bibfnamefont {K.~A.}\ \bibnamefont {Nelson}},\ }\bibfield  {title} {\enquote {\bibinfo {title} {Time-resolved vibrational spectroscopy in the impulsive limit},}\ }\bibfield  {booktitle} {\emph {\bibinfo {booktitle} {Chemical Reviews}},\ }\href {https://doi.org/10.1021/cr00025a006} {\bibfield  {journal} {\bibinfo  {journal} {Chemical Reviews}\ }\textbf {\bibinfo {volume} {94}},\ \bibinfo {pages} {157--193} (\bibinfo {year} {1994})}\BibitemShut {NoStop}%
\bibitem [{\citenamefont {Merlin}(1997)}]{Merlin:1997ca}%
  \BibitemOpen
  \bibfield  {author} {\bibinfo {author} {\bibfnamefont {R.}~\bibnamefont {Merlin}},\ }\bibfield  {title} {\enquote {\bibinfo {title} {Generating coherent thz phonons with light pulses},}\ }\bibfield  {booktitle} {\emph {\bibinfo {booktitle} {Highlights in Condensed Matter Physics and Materials Science}},\ }\href {https://doi.org/https://doi.org/10.1016/S0038-1098(96)00721-1} {\bibfield  {journal} {\bibinfo  {journal} {Solid State Communications}\ }\textbf {\bibinfo {volume} {102}},\ \bibinfo {pages} {207--220} (\bibinfo {year} {1997})}\BibitemShut {NoStop}%
\bibitem [{\citenamefont {Domingue}, \citenamefont {Winters},\ and\ \citenamefont {Bartels}(2014)}]{Domingue:2014bx}%
  \BibitemOpen
  \bibfield  {author} {\bibinfo {author} {\bibfnamefont {S.~R.}\ \bibnamefont {Domingue}}, \bibinfo {author} {\bibfnamefont {D.~G.}\ \bibnamefont {Winters}},\ and\ \bibinfo {author} {\bibfnamefont {R.~A.}\ \bibnamefont {Bartels}},\ }\bibfield  {title} {\enquote {\bibinfo {title} {Time-resolved coherent raman spectroscopy by high-speed pump-probe delay scanning},}\ }\bibfield  {booktitle} {\emph {\bibinfo {booktitle} {Optics Letters}},\ }\href {https://doi.org/10.1364/OL.39.004124} {\bibfield  {journal} {\bibinfo  {journal} {Optics Letters}\ }\textbf {\bibinfo {volume} {39}},\ \bibinfo {pages} {4124--4127} (\bibinfo {year} {2014})}\BibitemShut {NoStop}%
\bibitem [{\citenamefont {Raanan}\ \emph {et~al.}(2019)\citenamefont {Raanan}, \citenamefont {Audier}, \citenamefont {Shivkumar}, \citenamefont {Asher}, \citenamefont {Menahem}, \citenamefont {Yaffe}, \citenamefont {Forget}, \citenamefont {Rigneault},\ and\ \citenamefont {Oron}}]{Raanan2019}%
  \BibitemOpen
  \bibfield  {author} {\bibinfo {author} {\bibfnamefont {D.}~\bibnamefont {Raanan}}, \bibinfo {author} {\bibfnamefont {X.}~\bibnamefont {Audier}}, \bibinfo {author} {\bibfnamefont {S.}~\bibnamefont {Shivkumar}}, \bibinfo {author} {\bibfnamefont {M.}~\bibnamefont {Asher}}, \bibinfo {author} {\bibfnamefont {M.}~\bibnamefont {Menahem}}, \bibinfo {author} {\bibfnamefont {O.}~\bibnamefont {Yaffe}}, \bibinfo {author} {\bibfnamefont {N.}~\bibnamefont {Forget}}, \bibinfo {author} {\bibfnamefont {H.}~\bibnamefont {Rigneault}},\ and\ \bibinfo {author} {\bibfnamefont {D.}~\bibnamefont {Oron}},\ }\bibfield  {title} {\enquote {\bibinfo {title} {Sub-second hyper-spectral low-frequency vibrational imaging via impulsive raman excitation},}\ }\href@noop {} {\bibfield  {journal} {\bibinfo  {journal} {Optics letters}\ }\textbf {\bibinfo {volume} {44}},\ \bibinfo {pages} {5153--5156} (\bibinfo {year} {2019})}\BibitemShut {NoStop}%
\bibitem [{\citenamefont {Smith}\ \emph {et~al.}(2022)\citenamefont {Smith}, \citenamefont {Shivkumar}, \citenamefont {Field}, \citenamefont {Wilson}, \citenamefont {Rigneault},\ and\ \citenamefont {Bartels}}]{smith2022nearly}%
  \BibitemOpen
  \bibfield  {author} {\bibinfo {author} {\bibfnamefont {D.~R.}\ \bibnamefont {Smith}}, \bibinfo {author} {\bibfnamefont {S.}~\bibnamefont {Shivkumar}}, \bibinfo {author} {\bibfnamefont {J.}~\bibnamefont {Field}}, \bibinfo {author} {\bibfnamefont {J.~W.}\ \bibnamefont {Wilson}}, \bibinfo {author} {\bibfnamefont {H.}~\bibnamefont {Rigneault}},\ and\ \bibinfo {author} {\bibfnamefont {R.~A.}\ \bibnamefont {Bartels}},\ }\bibfield  {title} {\enquote {\bibinfo {title} {Nearly degenerate two-color impulsive coherent raman hyperspectral imaging},}\ }\href@noop {} {\bibfield  {journal} {\bibinfo  {journal} {Optics Letters}\ }\textbf {\bibinfo {volume} {47}},\ \bibinfo {pages} {5841--5844} (\bibinfo {year} {2022})}\BibitemShut {NoStop}%
\bibitem [{\citenamefont {Smith}\ \emph {et~al.}(2021)\citenamefont {Smith}, \citenamefont {Field}, \citenamefont {Winters}, \citenamefont {Domingue}, \citenamefont {Rininsland}, \citenamefont {Kane}, \citenamefont {Wilson},\ and\ \citenamefont {Bartels}}]{smith2021phase}%
  \BibitemOpen
  \bibfield  {author} {\bibinfo {author} {\bibfnamefont {D.~R.}\ \bibnamefont {Smith}}, \bibinfo {author} {\bibfnamefont {J.~J.}\ \bibnamefont {Field}}, \bibinfo {author} {\bibfnamefont {D.~G.}\ \bibnamefont {Winters}}, \bibinfo {author} {\bibfnamefont {S.~R.}\ \bibnamefont {Domingue}}, \bibinfo {author} {\bibfnamefont {F.}~\bibnamefont {Rininsland}}, \bibinfo {author} {\bibfnamefont {D.~J.}\ \bibnamefont {Kane}}, \bibinfo {author} {\bibfnamefont {J.~W.}\ \bibnamefont {Wilson}},\ and\ \bibinfo {author} {\bibfnamefont {R.~A.}\ \bibnamefont {Bartels}},\ }\bibfield  {title} {\enquote {\bibinfo {title} {Phase noise limited frequency shift impulsive raman spectroscopy},}\ }\href@noop {} {\bibfield  {journal} {\bibinfo  {journal} {APL Photonics}\ }\textbf {\bibinfo {volume} {6}},\ \bibinfo {pages} {026107} (\bibinfo {year} {2021})}\BibitemShut {NoStop}%
\bibitem [{\citenamefont {Raanan}\ \emph {et~al.}(2018)\citenamefont {Raanan}, \citenamefont {L{\"u}ttig}, \citenamefont {Silberberg},\ and\ \citenamefont {Oron}}]{Raanan:2018df}%
  \BibitemOpen
  \bibfield  {author} {\bibinfo {author} {\bibfnamefont {D.}~\bibnamefont {Raanan}}, \bibinfo {author} {\bibfnamefont {J.}~\bibnamefont {L{\"u}ttig}}, \bibinfo {author} {\bibfnamefont {Y.}~\bibnamefont {Silberberg}},\ and\ \bibinfo {author} {\bibfnamefont {D.}~\bibnamefont {Oron}},\ }\bibfield  {title} {\enquote {\bibinfo {title} {Vibrational spectroscopy via stimulated raman induced kerr lensing},}\ }\bibfield  {booktitle} {\emph {\bibinfo {booktitle} {APL Photonics}},\ }\href {https://doi.org/10.1063/1.5029809} {\bibfield  {journal} {\bibinfo  {journal} {APL Photonics}\ }\textbf {\bibinfo {volume} {3}},\ \bibinfo {pages} {092501} (\bibinfo {year} {2018})}\BibitemShut {NoStop}%
\bibitem [{\citenamefont {Wahlstrand}\ \emph {et~al.}(2005)\citenamefont {Wahlstrand}, \citenamefont {Merlin}, \citenamefont {Li}, \citenamefont {Cundiff},\ and\ \citenamefont {Martinez}}]{Wahlstrand:05}%
  \BibitemOpen
  \bibfield  {author} {\bibinfo {author} {\bibfnamefont {J.~K.}\ \bibnamefont {Wahlstrand}}, \bibinfo {author} {\bibfnamefont {R.}~\bibnamefont {Merlin}}, \bibinfo {author} {\bibfnamefont {X.}~\bibnamefont {Li}}, \bibinfo {author} {\bibfnamefont {S.~T.}\ \bibnamefont {Cundiff}},\ and\ \bibinfo {author} {\bibfnamefont {O.~E.}\ \bibnamefont {Martinez}},\ }\bibfield  {title} {\enquote {\bibinfo {title} {Impulsive stimulated raman scattering: comparison between phase-sensitive and spectrally filtered techniques},}\ }\href {https://doi.org/10.1364/OL.30.000926} {\bibfield  {journal} {\bibinfo  {journal} {Opt. Lett.}\ }\textbf {\bibinfo {volume} {30}},\ \bibinfo {pages} {926--928} (\bibinfo {year} {2005})}\BibitemShut {NoStop}%
\bibitem [{\citenamefont {Wilson}, \citenamefont {Schlup},\ and\ \citenamefont {Bartels}(2008{\natexlab{a}})}]{Wilson:2008lk}%
  \BibitemOpen
  \bibfield  {author} {\bibinfo {author} {\bibfnamefont {J.~W.}\ \bibnamefont {Wilson}}, \bibinfo {author} {\bibfnamefont {P.}~\bibnamefont {Schlup}},\ and\ \bibinfo {author} {\bibfnamefont {R.~A.}\ \bibnamefont {Bartels}},\ }\bibfield  {title} {\enquote {\bibinfo {title} {Synthetic temporal aperture coherent molecular phase spectroscopy},}\ }\href {https://doi.org/https://doi.org/10.1016/j.cplett.2008.08.067} {\bibfield  {journal} {\bibinfo  {journal} {Chemical Physics Letters}\ }\textbf {\bibinfo {volume} {463}},\ \bibinfo {pages} {300--304} (\bibinfo {year} {2008}{\natexlab{a}})}\BibitemShut {NoStop}%
\bibitem [{\citenamefont {Schlup}, \citenamefont {Wilson},\ and\ \citenamefont {Bartels}(2009)}]{Schlup:2009bq}%
  \BibitemOpen
  \bibfield  {author} {\bibinfo {author} {\bibfnamefont {P.}~\bibnamefont {Schlup}}, \bibinfo {author} {\bibfnamefont {J.~W.}\ \bibnamefont {Wilson}},\ and\ \bibinfo {author} {\bibfnamefont {R.~A.}\ \bibnamefont {Bartels}},\ }\bibfield  {title} {\enquote {\bibinfo {title} {Sensitive and selective detection of low-frequency vibrational modes through a phase-shifting fourier transform spectroscopy},}\ }\bibfield  {booktitle} {\emph {\bibinfo {booktitle} {IEEE Journal of Quantum Electronics}},\ }\href {https://doi.org/10.1109/JQE.2009.2013121} {\bibfield  {journal} {\bibinfo  {journal} {IEEE Journal of Quantum Electronics}\ }\textbf {\bibinfo {volume} {45}},\ \bibinfo {pages} {777--782} (\bibinfo {year} {2009})}\BibitemShut {NoStop}%
\bibitem [{\citenamefont {Wilson}, \citenamefont {Schlup},\ and\ \citenamefont {Bartels}(2008{\natexlab{b}})}]{Wilson:2008hh}%
  \BibitemOpen
  \bibfield  {author} {\bibinfo {author} {\bibfnamefont {J.~W.}\ \bibnamefont {Wilson}}, \bibinfo {author} {\bibfnamefont {P.}~\bibnamefont {Schlup}},\ and\ \bibinfo {author} {\bibfnamefont {R.}~\bibnamefont {Bartels}},\ }\bibfield  {title} {\enquote {\bibinfo {title} {Phase measurement of coherent raman vibrational spectroscopy with chirped spectral holography},}\ }\bibfield  {booktitle} {\emph {\bibinfo {booktitle} {Optics Letters}},\ }\href {https://doi.org/10.1364/OL.33.002116} {\bibfield  {journal} {\bibinfo  {journal} {Optics Letters}\ }\textbf {\bibinfo {volume} {33}},\ \bibinfo {pages} {2116--2118} (\bibinfo {year} {2008}{\natexlab{b}})}\BibitemShut {NoStop}%
\bibitem [{\citenamefont {van Dijk}, \citenamefont {Lippitz},\ and\ \citenamefont {Orrit}(2005)}]{van2005detection}%
  \BibitemOpen
  \bibfield  {author} {\bibinfo {author} {\bibfnamefont {M.~A.}\ \bibnamefont {van Dijk}}, \bibinfo {author} {\bibfnamefont {M.}~\bibnamefont {Lippitz}},\ and\ \bibinfo {author} {\bibfnamefont {M.}~\bibnamefont {Orrit}},\ }\bibfield  {title} {\enquote {\bibinfo {title} {Detection of acoustic oscillations of single gold nanospheres by time-resolved interferometry},}\ }\href@noop {} {\bibfield  {journal} {\bibinfo  {journal} {Physical review letters}\ }\textbf {\bibinfo {volume} {95}},\ \bibinfo {pages} {267406} (\bibinfo {year} {2005})}\BibitemShut {NoStop}%
\bibitem [{\citenamefont {Schlup}\ \emph {et~al.}(2007)\citenamefont {Schlup}, \citenamefont {Wilson}, \citenamefont {Hartinger},\ and\ \citenamefont {Bartels}}]{schlup2007dispersion}%
  \BibitemOpen
  \bibfield  {author} {\bibinfo {author} {\bibfnamefont {P.}~\bibnamefont {Schlup}}, \bibinfo {author} {\bibfnamefont {J.}~\bibnamefont {Wilson}}, \bibinfo {author} {\bibfnamefont {K.}~\bibnamefont {Hartinger}},\ and\ \bibinfo {author} {\bibfnamefont {R.~A.}\ \bibnamefont {Bartels}},\ }\bibfield  {title} {\enquote {\bibinfo {title} {Dispersion balancing of variable-delay monolithic pulse splitters},}\ }\href@noop {} {\bibfield  {journal} {\bibinfo  {journal} {Applied optics}\ }\textbf {\bibinfo {volume} {46}},\ \bibinfo {pages} {5967--5973} (\bibinfo {year} {2007})}\BibitemShut {NoStop}%
\bibitem [{\citenamefont {Ghosh}\ \emph {et~al.}(2021)\citenamefont {Ghosh}, \citenamefont {Herink}, \citenamefont {Perri}, \citenamefont {Preda}, \citenamefont {Manzoni}, \citenamefont {Polli},\ and\ \citenamefont {Cerullo}}]{ghosh2021broadband}%
  \BibitemOpen
  \bibfield  {author} {\bibinfo {author} {\bibfnamefont {S.}~\bibnamefont {Ghosh}}, \bibinfo {author} {\bibfnamefont {G.}~\bibnamefont {Herink}}, \bibinfo {author} {\bibfnamefont {A.}~\bibnamefont {Perri}}, \bibinfo {author} {\bibfnamefont {F.}~\bibnamefont {Preda}}, \bibinfo {author} {\bibfnamefont {C.}~\bibnamefont {Manzoni}}, \bibinfo {author} {\bibfnamefont {D.}~\bibnamefont {Polli}},\ and\ \bibinfo {author} {\bibfnamefont {G.}~\bibnamefont {Cerullo}},\ }\bibfield  {title} {\enquote {\bibinfo {title} {Broadband optical activity spectroscopy with interferometric fourier-transform balanced detection},}\ }\href@noop {} {\bibfield  {journal} {\bibinfo  {journal} {ACS photonics}\ }\textbf {\bibinfo {volume} {8}},\ \bibinfo {pages} {2234--2242} (\bibinfo {year} {2021})}\BibitemShut {NoStop}%
\bibitem [{\citenamefont {Bartels}\ \emph {et~al.}(2001)\citenamefont {Bartels}, \citenamefont {Weinacht}, \citenamefont {Wagner}, \citenamefont {Baertschy}, \citenamefont {Greene}, \citenamefont {Murnane},\ and\ \citenamefont {Kapteyn}}]{Bartels:2001wo}%
  \BibitemOpen
  \bibfield  {author} {\bibinfo {author} {\bibfnamefont {R.~A.}\ \bibnamefont {Bartels}}, \bibinfo {author} {\bibfnamefont {T.~C.}\ \bibnamefont {Weinacht}}, \bibinfo {author} {\bibfnamefont {N.}~\bibnamefont {Wagner}}, \bibinfo {author} {\bibfnamefont {M.}~\bibnamefont {Baertschy}}, \bibinfo {author} {\bibfnamefont {C.~H.}\ \bibnamefont {Greene}}, \bibinfo {author} {\bibfnamefont {M.~M.}\ \bibnamefont {Murnane}},\ and\ \bibinfo {author} {\bibfnamefont {H.~C.}\ \bibnamefont {Kapteyn}},\ }\bibfield  {title} {\enquote {\bibinfo {title} {Phase modulation of ultrashort light pulses using molecular rotational wave packets},}\ }\href {https://doi.org/10.1103/PhysRevLett.88.013903} {\bibfield  {journal} {\bibinfo  {journal} {Physical Review Letters}\ }\textbf {\bibinfo {volume} {88}},\ \bibinfo {pages} {013903--} (\bibinfo {year} {2001})}\BibitemShut {NoStop}%
\bibitem [{\citenamefont {Gershgoren}\ \emph {et~al.}(2003)\citenamefont {Gershgoren}, \citenamefont {Bartels}, \citenamefont {Fourkas}, \citenamefont {Tobey}, \citenamefont {Murnane},\ and\ \citenamefont {Kapteyn}}]{Gershgoren:2003tx}%
  \BibitemOpen
  \bibfield  {author} {\bibinfo {author} {\bibfnamefont {E.}~\bibnamefont {Gershgoren}}, \bibinfo {author} {\bibfnamefont {R.~A.}\ \bibnamefont {Bartels}}, \bibinfo {author} {\bibfnamefont {J.~T.}\ \bibnamefont {Fourkas}}, \bibinfo {author} {\bibfnamefont {R.}~\bibnamefont {Tobey}}, \bibinfo {author} {\bibfnamefont {M.~M.}\ \bibnamefont {Murnane}},\ and\ \bibinfo {author} {\bibfnamefont {H.~C.}\ \bibnamefont {Kapteyn}},\ }\bibfield  {title} {\enquote {\bibinfo {title} {Simplified setup for high-resolution spectroscopy that uses ultrashort pulses},}\ }\bibfield  {booktitle} {\emph {\bibinfo {booktitle} {Optics Letters}},\ }\href {https://doi.org/10.1364/OL.28.000361} {\bibfield  {journal} {\bibinfo  {journal} {Optics Letters}\ }\textbf {\bibinfo {volume} {28}},\ \bibinfo {pages} {361--363} (\bibinfo {year} {2003})}\BibitemShut {NoStop}%
\bibitem [{\citenamefont {Wilson}\ and\ \citenamefont {Bartels}(2011)}]{wilson2011rapid}%
  \BibitemOpen
  \bibfield  {author} {\bibinfo {author} {\bibfnamefont {J.~W.}\ \bibnamefont {Wilson}}\ and\ \bibinfo {author} {\bibfnamefont {R.~A.}\ \bibnamefont {Bartels}},\ }\bibfield  {title} {\enquote {\bibinfo {title} {Rapid birefringent delay scanning for coherent multiphoton impulsive raman pump--probe spectroscopy},}\ }\href@noop {} {\bibfield  {journal} {\bibinfo  {journal} {IEEE Journal of Selected Topics in Quantum Electronics}\ }\textbf {\bibinfo {volume} {18}},\ \bibinfo {pages} {130--139} (\bibinfo {year} {2011})}\BibitemShut {NoStop}%
\bibitem [{\citenamefont {Bartels}\ \emph {et~al.}(2002)\citenamefont {Bartels}, \citenamefont {Weinacht}, \citenamefont {Leone}, \citenamefont {Kapteyn},\ and\ \citenamefont {Murnane}}]{Bartels:2002le}%
  \BibitemOpen
  \bibfield  {author} {\bibinfo {author} {\bibfnamefont {R.~A.}\ \bibnamefont {Bartels}}, \bibinfo {author} {\bibfnamefont {T.~C.}\ \bibnamefont {Weinacht}}, \bibinfo {author} {\bibfnamefont {S.~R.}\ \bibnamefont {Leone}}, \bibinfo {author} {\bibfnamefont {H.~C.}\ \bibnamefont {Kapteyn}},\ and\ \bibinfo {author} {\bibfnamefont {M.~M.}\ \bibnamefont {Murnane}},\ }\bibfield  {title} {\enquote {\bibinfo {title} {Nonresonant control of multimode molecular wave packets at room temperature},}\ }\href {https://doi.org/10.1103/PhysRevLett.88.033001} {\bibfield  {journal} {\bibinfo  {journal} {Physical Review Letters}\ }\textbf {\bibinfo {volume} {88}},\ \bibinfo {pages} {033001--} (\bibinfo {year} {2002})}\BibitemShut {NoStop}%
\bibitem [{\citenamefont {Ren}\ \emph {et~al.}(2019)\citenamefont {Ren}, \citenamefont {Hurwitz}, \citenamefont {Raanan}, \citenamefont {Oulevey}, \citenamefont {Oron},\ and\ \citenamefont {Silberberg}}]{Ren:19}%
  \BibitemOpen
  \bibfield  {author} {\bibinfo {author} {\bibfnamefont {L.}~\bibnamefont {Ren}}, \bibinfo {author} {\bibfnamefont {I.}~\bibnamefont {Hurwitz}}, \bibinfo {author} {\bibfnamefont {D.}~\bibnamefont {Raanan}}, \bibinfo {author} {\bibfnamefont {P.}~\bibnamefont {Oulevey}}, \bibinfo {author} {\bibfnamefont {D.}~\bibnamefont {Oron}},\ and\ \bibinfo {author} {\bibfnamefont {Y.}~\bibnamefont {Silberberg}},\ }\bibfield  {title} {\enquote {\bibinfo {title} {Terahertz coherent anti-stokes raman scattering microscopy},}\ }\href {https://doi.org/10.1364/OPTICA.6.000052} {\bibfield  {journal} {\bibinfo  {journal} {Optica}\ }\textbf {\bibinfo {volume} {6}},\ \bibinfo {pages} {52--55} (\bibinfo {year} {2019})}\BibitemShut {NoStop}%
\bibitem [{\citenamefont {Hartinger}\ and\ \citenamefont {Bartels}(2008)}]{Hartinger:2008oe}%
  \BibitemOpen
  \bibfield  {author} {\bibinfo {author} {\bibfnamefont {K.}~\bibnamefont {Hartinger}}\ and\ \bibinfo {author} {\bibfnamefont {R.~A.}\ \bibnamefont {Bartels}},\ }\bibfield  {title} {\enquote {\bibinfo {title} {Single-shot measurement of ultrafast time-varying phase modulation induced by femtosecond laser pulses with arbitrary polarization},}\ }\bibfield  {booktitle} {\emph {\bibinfo {booktitle} {Applied Physics Letters}},\ }\href {https://doi.org/10.1063/1.2801515} {\bibfield  {journal} {\bibinfo  {journal} {Applied Physics Letters}\ }\textbf {\bibinfo {volume} {92}},\ \bibinfo {pages} {021126} (\bibinfo {year} {2008})}\BibitemShut {NoStop}%
\bibitem [{\citenamefont {Domingue}\ and\ \citenamefont {Bartels}(2015)}]{domingue2015nearly}%
  \BibitemOpen
  \bibfield  {author} {\bibinfo {author} {\bibfnamefont {S.}~\bibnamefont {Domingue}}\ and\ \bibinfo {author} {\bibfnamefont {R.}~\bibnamefont {Bartels}},\ }\bibfield  {title} {\enquote {\bibinfo {title} {Nearly transform-limited sub-20-fs pulses at 1065 nm and> 10 nj enabled by a flat field ultrafast pulse shaper},}\ }\href@noop {} {\bibfield  {journal} {\bibinfo  {journal} {Optics Letters}\ }\textbf {\bibinfo {volume} {40}},\ \bibinfo {pages} {253--256} (\bibinfo {year} {2015})}\BibitemShut {NoStop}%
\bibitem [{\citenamefont {Wilson}\ \emph {et~al.}(2008)\citenamefont {Wilson}, \citenamefont {Schlup}, \citenamefont {Lunacek}, \citenamefont {Whitley},\ and\ \citenamefont {Bartels}}]{JesseCalibrattion}%
  \BibitemOpen
  \bibfield  {author} {\bibinfo {author} {\bibfnamefont {J.~W.}\ \bibnamefont {Wilson}}, \bibinfo {author} {\bibfnamefont {P.}~\bibnamefont {Schlup}}, \bibinfo {author} {\bibfnamefont {M.}~\bibnamefont {Lunacek}}, \bibinfo {author} {\bibfnamefont {D.}~\bibnamefont {Whitley}},\ and\ \bibinfo {author} {\bibfnamefont {R.~A.}\ \bibnamefont {Bartels}},\ }\bibfield  {title} {\enquote {\bibinfo {title} {Calibration of liquid crystal ultrafast pulse shaper with common-path spectral interferometry and application to coherent control with a covariance matrix adaptation evolutionary strategy},}\ }\bibfield  {booktitle} {\emph {\bibinfo {booktitle} {Review of Scientific Instruments}},\ }\href {https://doi.org/10.1063/1.2839919} {\bibfield  {journal} {\bibinfo  {journal} {Review of Scientific Instruments}\ }\textbf {\bibinfo {volume} {79}},\ \bibinfo {pages} {033103} (\bibinfo {year} {2008})}\BibitemShut {NoStop}%
\bibitem [{\citenamefont {Sinjab}\ \emph {et~al.}(2020)\citenamefont {Sinjab}, \citenamefont {Hashimoto}, \citenamefont {Zhao}, \citenamefont {Nagashima},\ and\ \citenamefont {Ideguchi}}]{Sinjab:20}%
  \BibitemOpen
  \bibfield  {author} {\bibinfo {author} {\bibfnamefont {F.}~\bibnamefont {Sinjab}}, \bibinfo {author} {\bibfnamefont {K.}~\bibnamefont {Hashimoto}}, \bibinfo {author} {\bibfnamefont {X.}~\bibnamefont {Zhao}}, \bibinfo {author} {\bibfnamefont {Y.}~\bibnamefont {Nagashima}},\ and\ \bibinfo {author} {\bibfnamefont {T.}~\bibnamefont {Ideguchi}},\ }\bibfield  {title} {\enquote {\bibinfo {title} {Enhanced spectral resolution for broadband coherent anti-stokes raman spectroscopy},}\ }\href {https://doi.org/10.1364/OL.388624} {\bibfield  {journal} {\bibinfo  {journal} {Opt. Lett.}\ }\textbf {\bibinfo {volume} {45}},\ \bibinfo {pages} {1515--1518} (\bibinfo {year} {2020})}\BibitemShut {NoStop}%
\end{thebibliography}%

\end{document}